# Piezoresistive Properties of Suspended Graphene Membranes under Uniaxial and Biaxial Strain in Nanoelectromechanical Pressure Sensors

Anderson D. Smith[1], Frank Niklaus[1], Alan Paussa[2], Stephan Schröder[1], Andreas C. Fischer[1], Mikael Sterner[1], Stefan Wagner[3], Sam Vaziri[1], Fredrik Forsberg[1], David Esseni[2], Mikael Ostling[1], and Max C. Lemme[1,3]

[1]KTH Royal Institute of Technology, Stockholm, Sweden
[2]DIEGM, University of Udine, Italy
[3]University of Siegen, Department of Electrical Engineering and Computer Science, Germany

*Address correspondence to: frank.niklaus@ee.kth.se, max.lemme@uni-siegen.de

## Abstract

Graphene membranes act as highly sensitive transducers in nanoelectromechanical devices due to their ultimate thinness. Previously, the piezoresistive effect has been experimentally verified in graphene using uniaxial strain in graphene. Here we report experimental and theoretical data on the uni- and biaxial piezoresistive properties of suspended graphene membranes applied to piezoresistive pressure sensors. A detailed model that utilizes a linearized Boltzman transport equation describes accurately the charge carrier density and mobility in strained graphene, and hence the gauge factor. The gauge factor is found to be practically independent of the doping concentration and crystallographic orientation of the graphene films. These investigations provide deeper insight into the piezoresistive behavior of graphene membranes.

Keywords: graphene, nanoelectromechanical system, NEMS, MEMS, strain gauge, transducer, piezoresistive transduction, gauge factor, pressure transducer, (suspended) graphene membranes, uniaxial and biaxial strain

Graphene, with its high carrier mobility,[1,2] low mass and high Young's modulus of over 1 TPa for both exfoliated and CVD graphene[3,4] shows great potential for future nanoelectromechanical system (NEMS) applications. Furthermore, graphene is stretchable up to 20% and allows for elastic recovery.[5] It adheres strongly to $SiO_2$ substrates despite the fact that the bonding is primarily based on van der Waals interactions.[6] Graphene is further impermeable to gasses, including helium,[7] and is therefore well suited for pressure sensor applications. In previous work, we have introduced pressure sensors based on suspended graphene membranes with direct electrical readout utilizing the piezoresistive effect in graphene.[8] In these sensors, a pressure difference between the two sides of a suspended graphene membrane deflects and strains the graphene. The rectangular shape of the membrane with a high width to length ratio leads to uniaxial strain. Due to the extreme thinness of the graphene membrane, the sensitivity normalized by the membrane area of these sensors is significantly higher than in conventional pressure sensors utilizing *e.g.* silicon membranes. Extracted gauge factors of approximately 2-3 for uniaxial strain in graphene membranes match well with theory and experimental results reported in literature.[9–12] Preliminary results from experiments with membrane geometries resulting in biaxially strained graphene indicate similar gauge factors.[13]

This paper extends in experiment and theory the exploration of graphene's mechanical properties for use in high-sensitivity piezoresistive nanoelectromechanical sensing. Simulations predict that strain affects the electronic structure of graphene with a dependence on the strain type (*i.e.* uniaxial, biaxial or shear strain).[14–17] We measured different sensor geometries and performed corresponding transport calculations using the linearized Boltzmann transport equation (LBTE) to investigate the piezoresistive effect in graphene due to uniaxial and biaxial (radial) strain. We present an improved model that includes the charge carrier density in graphene to quantitatively match the experiment. We establish that the gauge factor is independent of the doping concentration. Both our experiments as well as the simulations corroborate previous results and also suggest that graphene's piezoresistive behavior is independent of crystallographic orientation. The piezoresistive behavior can be described as a superposition of carrier density and carrier mobility modification, where the latter dominates. We further demonstrate that for small strain, graphene's intrinsic gauge factor is largely invariant of strain magnitude.

**Results**

Graphene pressure sensors were fabricated from chemical vapor deposited (CVD) graphene transferred from a copper foil onto an oxidized silicon substrate. The silicon dioxide ($SiO_2$) layer is 1.5 µm with thick, with etched cavities that are 1.5 µm deep. Upon transfer, the graphene covers the cavities and seals air at ambient pressure inside. The graphene is contacted through gold electrodes that are embedded in the $SiO_2$ layer prior to graphene transfer and is etched into patches using oxygen plasma. The graphene membranes suspended over the cavities make up the active part of the graphene patch. Fig. 1a shows a schematic of the device, Fig. 1b shows the device concept and Fig. 1c depicts the fabrication process, with details of the device fabrication described in the methods section.

When these suspended graphene membranes are placed inside a pressure chamber, the difference of the pressure in the sealed cavity and in the pressure chamber causes the graphene membrane to deflect – thereby applying strain (Fig. 1b). Devices without cavities have been used for control measurements and they have been shown to have negligible sensitivity to pressure changes[8] (see supporting information). Fig. 1d shows two color enhanced scanning electron microscope (SEM) images of fabricated devices that have been wire bonded and packaged. The graphene is shaded in blue, the cavities in green, the metal electrodes and contact pads in yellow, and the bond wires in orange. The resulting strain distribution in the graphene membranes is defined by the membrane shape (*i.e.* the cavity shape). The top SEM in Fig. 1d shows a rectangular graphene membrane, which results in a near uniaxial strain distribution in the membrane. The bottom SEM in Fig. 1d shows a circular graphene membrane, which results in a biaxial (radially symmetric) strain distribution in the membrane. To the right of each image are higher resolution SEM images with the respective cavity regions.

The strain levels in the graphene membranes depend on the applied pressure difference. The strain distribution and the strain levels affect the electronic properties of the membrane, which can be quantified by measuring the resistance of the graphene as it is strained. In the experiments, the sensors were placed in a pressure chamber. Prior to the measurements the pressure chamber was evacuated and purged repeatedly with inert argon gas to mitigate humidity and gas sensing effects in the graphene patch.[8,18] The chamber pressure was varied between 1000 mbar and a minimum of 100 mbar, the latter to reduce the risk of rupture of the membrane or delamination of the graphene from the $SiO_2$ surface at the membrane edge. Delamination has been shown to occur at strains on the order of MPa and so can be reasonably neglected for the strain levels examined in this work.[6] A Wheatstone bridge was used to measure the resistance of the graphene, with the graphene patch acting as the variable resistor. Before each measurement, the resistance of the bridge was calibrated with a potentiometer after the pressure chamber was filled with Ar to atmospheric pressure. Device self-heating effects from the biasing current were kept to a minimum by limiting the bias of the Wheatstone bridge to 200 mV square wave pulses with a length of 500 µs. The voltage output signal of the Wheatstone bridge was then amplified, passed through a low pass filter, sampled using an analog-to-digital converter and finally converted into a corresponding resistance value. The graphene membrane devices were compared to a control group of devices that contained graphene patches without cavities (*i.e.* no suspended membranes) (see supporting information). A capacitor model has been applied to analyze the potential effect of capacitive coupling between the suspended graphene membrane and fixed charges in the substrate. The model predicts the resistance change in the graphene caused by capacitive coupling with fixed charges in the substrate. The analysis indicates that the effect of capacitive coupling on the device resistance is very small in comparison to the piezoresistive effect (see Supporting Information).

Fig. 2a compares the resistance response of a circular graphene pressure sensor with a diameter of 18 µm over time with the readout of a commercial digital vacuum transducer PDR 900 (MKS Instruments). The graph further includes the average strain (in percent) in the graphene membrane obtained by COMSOL modeling of the graphene deflection. This model has been found to be reasonably accurate when compared to AFM measurements of membrane deflection reported in literature[6–8]. The strain-deflection COMSOL model parameters are outlined in the Methods Section. The graphene patch resistance was measured repeatedly for the same 18 µm diameter membrane and is plotted as a function of the calculated average strain in Fig. 2b (blue circles). The data can be fitted linearly (red line) in most measurement cycles, which matches well with the LBTE simulations. This is significant because it will allow for simple sensor calibration, even though in some cases the curves deviate from being linear. We attribute this to parasitic effects the nature of which cannot be identified accurately in the experiment.

The data from Fig. 2b was used to calculate the change of resistance in percent (ΔR/R) of only the suspended part of the graphene, *i.e.* the nanoelectromechanically active membrane area (Fig. 2c). An equivalent resistor network model of the graphene patch was implemented (Fig. 2d). Fig. 2d shows close up SEM images of the cavity regions of a 24 µm circular membrane and a 6 µm x 64 µm rectangular membrane. The colored regions indicate different resistor components used in the model. The variable resistance of the suspended graphene membrane is represented by $R_2$, displayed in yellow. For the circular membrane devices, the resistance of the membrane region $R_2$ is approximated by a square region of equal area (Fig. 2d-1) for simplicity. $R_2$ can be extracted through Eq. 1, with the assumption that only the resistance in the cavity region changes with strain, while $R_1$, $R_3$, $R_4$, and $R_5$ remain constant and $R_{tot}$ is the measured resistance of the graphene device as a whole. Reference measurements of graphene patches with identical sizes as the sensor devices, but without cavities showed no change in the resistance as a function of pressure.

$$R_{tot} = R_1 + \frac{1}{\frac{1}{R_2}+\frac{1}{R_4}+\frac{1}{R_5}} + R_3 \qquad (1)$$

The resistor model was applied to three different membrane shapes and the relative resistance change is plotted *versus* the average strain obtained from LBTE calculations (Fig. 2c). The experimental data in Fig. 2c stems from circular membranes with a diameter of 24 µm and 18 µm (biaxial strain, blue diamonds and grey x's, respectively) and from rectangular membranes with a size of 6 µm x 64 µm (uniaxial strain, orange dots). One 24 µm device, one 18 µm devices and two 6 µm x 64 µm devices have been investigated along with 3 control devices which do not have a cavity[8] (see supporting information). The latter ensures that the presence of a cavity is the driving force behind the device's pressure sensitivity. The percent change in resistance for all devices was normalized with respect to the lowest experimentally measured resistance. The linear fits to each measurement also reflect their respective normalization.

Charge transport simulations have been performed to estimate the expected percentage resistance change in the graphene membrane in the case of both biaxial and uniaxial strain and are described in detail in the methods section. The results from these simulations are included in Fig.2c as red (biaxial strain) and blue (uniaxial strain) dots. The calculated values are somewhat lower than the experimental data. Fig. 2e displays the extracted gauge factors *versus* chamber pressure for the three devices and the simulations from Fig. 2c. For each device type, the gauge factor is calculated by dividing the measured change in resistance of the membrane area by the calculated average strain in the membrane. Each device data set has been linearly fitted and the linear fits of the gauge factors remain nearly constant for all pressure levels and, as a consequence, all strains (see supporting information). The gauge factors for the two circular membranes with biaxial (radial) strain are approximately equal. The experimental gauge factors exceed the simulations. Although this discrepancy is not fully understood, we attribute it to possible local defects and grain boundaries present in the CVD graphene that have been shown to result in higher gauge factors[19] and that are not present in the model. Another important outcome of this study is that the gauge factor is independent of the applied strain for all measured devices. Furthermore, the gauge factors of the two circular membranes are very similar, *i.e.* nearly independent of the cavity diameter. This is consistent with LBTE simulation (see Figure 2d). However, our simulations (see Fig. 2d) as well as some reports in literature (see Table 1) suggest that that the gauge factor in uniaxially strained graphene (*i.e.* rectangular membranes) should be slightly higher than in biaxially strained graphene (*i.e.* circular membranes). The source for this discrepancy is not clear at this point, but may be a combination of other sources of influence, *i.e.* slight temperature variations, residual humidity, random grain boundaries or capacitive coupling (see Supporting Information). Further, as the dominant mechanism in graphene's piezoresistivity is thought to arise from neutral defect scattering, there may be a significant contribution from pseudo-spin flips induced by scattering centers as proposed by Couto *et al.*[20] These additional phenomena have not been considered in our model, may explain the discrepancy and provides an interesting motivation for future investigation.

We report simulated gauge factors of 1.25 for biaxial strain and 2.20 for uniaxial strain. For experimental data, we report gauge factors in graphene membranes of 6.73 for biaxial strain (circular membranes) and 3.91 for uniaxial strain (rectangular membranes). These values are in the same range as values reported in literature obtained with different methods: Zhu *et. al.* reports a gauge factor of 1.6[9] (biaxial), Huang reports 1.9[10] (uniaxial), Lee reports 6.1[11] (uniaxial) and Wang reports a gauge factor of 2[12] (uniaxial), as summarized in Table 1. Our values are consistent with experimental and simulated literature data, suggesting that the gauge factor of graphene very likely resides in the range from 1.5 to 7.

The following theoretical analysis discusses in detail the physical origin of the observed gauge factors – specifically how strain affects the charge transport in the graphene membrane. The strain affects both carrier mobility and carrier density, as shown Fig. 3a, which compares the magnitude of change in charge density with that of the carrier mobility. From Fig. 3a, the strain affects mobility and carrier density in opposite directions with the latter partially compensating the former. Details of the models are described in the methods section. The central message of the calculations is a resistivity that is approximately proportional to the inverse of the squared Fermi velocity:

$$\rho_e(\epsilon) \sim \frac{1}{v_F(\epsilon)^2} \qquad (2)$$

The resistivity increases with strain, since the Fermi velocity decreases with strain.

An important feature of the piezoresistivity of graphene is its independence to changes in crystallographic orientation for small uniaxial strain.[8] This allows the possibility of using randomly oriented membranes, such as one would expect from CVD graphene, for strain sensor applications. Here, we also take into account the strain distribution in a circular membrane, including the membrane's radius. In a circular membrane, the strain levels are at a maximum at its center and the carrier mobility is at a minimum. We have calculated the average mobility as a function of membrane radius for three crystallographic orientations, namely armchair (60˚), zigzag (30˚), and 45˚ (Fig 3b). Previously, our simulations assumed the carrier density in the graphene membrane to be intrinsic (n = $10^{11}$ cm$^{-2}$). However, CVD graphene typically exhibits a much larger carrier density than graphene under perfect conditions due to electrostatic doping effects, *e.g.* through the substrate. Therefore, a case of much higher carrier concentration of n ~ 3 x $10^{12}$ cm$^{-2}$ was simulated with identical conditions, *i.e.* radial strain distribution with the largest strain in the center of the membrane and strain directions of armchair (60˚), zigzag (30˚), and 45˚. The value of the carrier concentration was chosen based on extraction from more than 4000 graphene transistor devices fabricated in the same process flow as the sensors. Further, this charge density value conforms well with previous literature.[21,22] We find that the influence of the crystallographic orientation on the device behavior is negligible, whereas the radius (and therefore the strain magnitude) has quite a strong influence (Fig. 3d).

The strong influence of the radius on strain and resistivity is observed also for highly doped graphene membranes (Fig. 4a), while the gauge factor is again independent of the direction of applied strain (inset of Fig 4a), as is the gauge factor (Fig 4b). This is important because our analysis suggests that neither large variations in the doping concentration nor variations in the crystallographic orientation of the graphene are expected to significantly affect the gauge factor.

So far, as in previous studies,[8,13] gauge factors have been extracted using a simplified resistor network model (see above) in combination with a COMSOL model that calculates the strain in a graphene membrane for a given deflection (due to pressure differences acting on the device). Here, a refined finite element model is introduced. This model incorporates LBTE simulations to calculate the charge carrier density and mobility for a given strain at each element of the mesh defined over the graphene membrane. In addition, the model is used to simulate the entire patch region of the graphene – not just the membrane area. This allows the calculation of the total patch resistance and its comparison to experimental data. The improved model further includes doping effects, as previously discussed. LBTE simulations predict a linear relationship between strain and resistivity for small strain. The predicted resistance of an entire graphene patch with an 18 μm diameter circular membrane is approximately 1.05 kΩ, close to the measured value of 1.1 kΩ. In the improved model, a charge carrier density of $n_0 = 3.5 \cdot 10^{12}$ cm$^{-2}$ and a mobility of 1000 cm$^2$V$^{-1}$s$^{-1}$ were assumed. Both values were chosen based on extracted values from 4500 graphene transistors fabricated using the same process flow as the sensors and using an established extraction method.[23] The values further agree well with previous literature for both the carrier density[21,22] as well as the mobility.[24–26] The spatial distribution of the strain induced resistivity change of the corresponding graphene patch is shown in Fig. 4c. Using this model for the entire graphene patch, the simplified resistor model was employed to extract gauge factors, from the refined model, and the resulting gauge factors were compared to the LBTE based model. Even though the simplified model cannot be used to predict the resistivity of a graphene patch, it can be used to calculate accurately the gauge factor of the sensors from electrical measurements (Fig. 4d). The refined model, in contrast, allows gauge factor predictions on its own. Through comparison of the LBTE gauge factor extraction to the simplified model extraction, we confirm that the simplified model provides a high precision approximation of the device's gauge factor.

## Conclusions

Graphene behaves somewhat differently depending on whether it is strained biaxially or uniaxially. The gauge factors extracted in this work fall within the range of values reported by previous literature and remain unchanging along different pressure ranges, as predicted by theory. Although simulation predicts that biaxial strain will produce slightly lower gauge factors than uniaxial strain, in our experiments the biaxial strained devices consistently had higher gauge factors than uniaxially strained, which suggests that there may be mechanisms at work which are not fully understood. Further, simulations suggest that neither the crystallographic orientation of the graphene, nor the doping concentration significantly affects the gauge factor for the strains examined. Also, the validity of device gauge factor extraction using a simplified resistance model approximation is affirmed.

## Methods

Devices were fabricated on pre-doped silicon substrate with p-type phosphorous doping. A layer of silicon dioxide ($SiO_2$) was thermally grown (Fig. 1c-1). Next, rectangular and circular cavities of different sizes were etched into the $SiO_2$ layer at a depth of 1.5 µm using a resist mask and reactive ion etching (RIE) with Ar and $CHF_3$ as etch gasses at an electromagnetic field power of 200 W and a gas pressure of 40 mTorr. Fig. 1c-2 shows an example of a circular cavity with a diameter of 24 µm. Long rectangular cavities result in near uniaxial strain in the graphene membrane. In contrast, circular cavities result in biaxial strain in the graphene membrane spanning over the cavity, with a radial gradient. RIE was used to etch 640 nm deep holes into the $SiO_2$ layer for contact electrodes and bond pads. These holes were then filled with a 160 nm thick layer of Ti to act as an adhesion layer and a 500 nm thick contact layer of gold (Fig. 1c-3). The purpose of realizing the contact electrodes before the graphene transfer is to limit the number of fabrication steps after the graphene is transferred, as the membranes are extremely fragile. Further, photoresist residue is known to contaminate the surface of graphene when using standard lithography processes after graphene transfer. By placing the metal electrodes underneath the graphene, we ensure that there is no residue between the electrodes and the graphene layer. Next, graphene is transferred to the pre-processed substrate (Fig. 1c-4). Chemical vapor deposited (CVD) graphene grown on copper foil from a commercial vendor (www.graphenea.com) was used. A layer of poly(Bisphenol A) carbonate (PC) was spin coated onto the surface of the graphene. This polymer film acts as a carrier to transport the graphene from the initial copper foil to the target substrate.[26–30] Carbon residues on the backside of the copper foil were then etched using $O_2$ plasma. The exposed side of the copper foil was placed into a solution of $FeCl_3$ to etch the copper foil– leaving a free floating graphene membrane on the polymer sheet. The graphene was then fished out of the solution using the target substrate and placed into HCl and DI water in order to clean away Fe ions. Next, the substrate/graphene stack was placed on a hotplate at 45 °C for 10 minutes. This was done in order to improve the adhesion between the graphene and the substrate. After heating, the substrate was allowed to cool before leaving it in a solution of chloroform for about 18 hours to remove the polymer layer on the graphene. The graphene was selectively etched using $O_2$ plasma in combination with a standard photolithography mask (Fig. 1c-5). Once the devices were completed, they were wire bonded into a ceramic chip package to facilitate measurements in a vacuum chamber. Fig. 1c-6 shows a photograph of the chip package and Fig. 1d shows two color-enhanced SEMs, one of a sensor with a rectangular graphene membrane and another one of a sensor with a circular graphene membrane. Raman Spectroscopy and electrical measurements were performed on similarly transferred devices to verify the presence of graphene (see Supporting Information).

Simulations of the strain and deflection for different membrane areas were performed in COMSOL Multiphysics and were found to match well with literature data.[8,32] The model is comprised of isotropic, linearly elastic 2D plates, which are clamped along the cavity boundary. The material parameters used in the model are as follows: Poisson's ratio *v* = 0.16 and elastic constant $E_t$ = 347 N/m (based on Koenig et al.)[6] and *t* is the membrane thickness (*t* = 0.335 nm).

Charge transport simulations were performed using the LBTE approach, which assumes a vanishing electric field in the transport direction, and either for an intrinsic graphene condition, corresponding to a Fermi level at Dirac point ($E_F = 0$eV), or for a graphene electron concentration of about $3.1 \times 10^{12}$ [cm$^{-2}$], corresponding to $E_F \sim 0.175 m$eV. The resistance of a given graphene membrane region is then defined by Eq. 3.

$$R_2 = \rho \frac{L'}{W_t'} = \frac{1}{2qN_e\mu_e} \frac{(1+\varepsilon_{xx})L}{(1+\varepsilon_{yy})W_t}, \quad (3)$$

Where $\rho$ respresents the resistivity of the membrane region, $N_e$ is the electron density, and $\mu_e$ is the electron mobility. At the Dirac point, both the electron and hole density and mobility are the same. $\varepsilon_{xx}$

and $\varepsilon_{yy}$ are strain components in the x and y directions of the membrane region. Furthermore, $q$ is the fundamental unit of charge, and $L'$ and $W_t'$ are the length and width change of the device as it is strained. As can be seen by Eq. 3, the resistance of the graphene membrane $R_2$ can change due to a change in several factors: charge density, carrier mobility, and device geometry as a result of a deflection of the graphene membrane. None of these possible effects can be neglected *a priori*. Since anisotropic strain leads to anisotropic energy dispersion relations, we have employed a recently developed general approach for the deterministic solution of the Linearized Boltzmann Transport Equation[33] and we used it to relate the effect of strain on the electron mobility $\mu_e$. This approach can naturally handle anisotropic energy dispersion relations and scattering rates, allowing us to accurately estimate the dependence of the resistance $R_2$ on the variation of the strain tensor or on the alignment between the transport direction and the graphene orientation. In order to gain an intuitive insight in the results of numerical simulations, we here develop a simplified analysis and assume that the shear strain components are negligible and the strain tensor takes the simple form of Eq. 4.

$$\begin{pmatrix} \epsilon & 0 \\ 0 & \epsilon \end{pmatrix} \qquad (4)$$

The energy dispersion relation E(k) near the Dirac energy remains linear and isotropic also in the presence of strain. Hence, E(k) can be still expressed by Eq. 5 where $v_F(\epsilon)$ is the Fermi velocity, that depends on the strain.

$$E(k) = h v_F(\epsilon) k \qquad (5)$$

Fermi velocity will decrease with the strain, which implies that for a given energy the magnitude k of the wave vector is larger for strained graphene than in relaxed graphene. As a consequence, the density of states increases with the strain, because it is approximately proportional to the inverse of the squared Fermi velocity:

$$N_e(\epsilon) \sim \frac{1}{v_F(\epsilon)^2} \qquad (6)$$

An increase of the carrier density leads to a decrease of the resistance $R_2$ (see Eq.2), and this behavior is not consistent with the results from our experiments. Hence, a simple charge based analysis does not explain the dependence of the resistance on the strain. For this reason we have to analyze also the dependence of the mobility on the strain. If we suppose that the neutral defects[34] are the predominant scattering mechanism in CVD graphene[35] and if we model their scattering rate with the expression presented in,[34] it is possible to demonstrate that the dependence of the mobility on the Fermi velocity is given by Eq. 7.

$$\mu_e(\epsilon) \sim v_F(\epsilon)^4 \qquad (7)$$

The mobility therefore decreases with the strain, leading to an increase of the resistor $R_2$ (see Eq. 2).

## Supporting Information

In the supporting information, we provide a detailed explanation of the measurement setup. We also present details of the analysis of the effect of capacitive coupling on the device behavior and a comparison to results reported in literature, confirming that capacitive coupling cannot account for the measured resistance change in the graphene membrane. In addition, the relationship between the membrane shape and the type of strain it produces is verified using finite element models. Further, the relationship between gauge factor and strain is discussed in greater detail. Finally, a comparison of the devices to a control device with no suspended membrane is demonstrated.

## Acknowledgements

Support from the European Commission through two ERC Starting Grants (InteGraDe, 307311 & M&M's, 277879), the Swedish Research Council (E0616001 and D0575901, iGRAPHENE), the German Research Foundation (DFG, LE 2440/1-2) and the German Federal Ministry of Education and Research (BMBF, 03XP0006C) is gratefully acknowledged. The authors thank Christian Wendlandt for fruitful discussions about the structural mechanics modeling of the membranes.

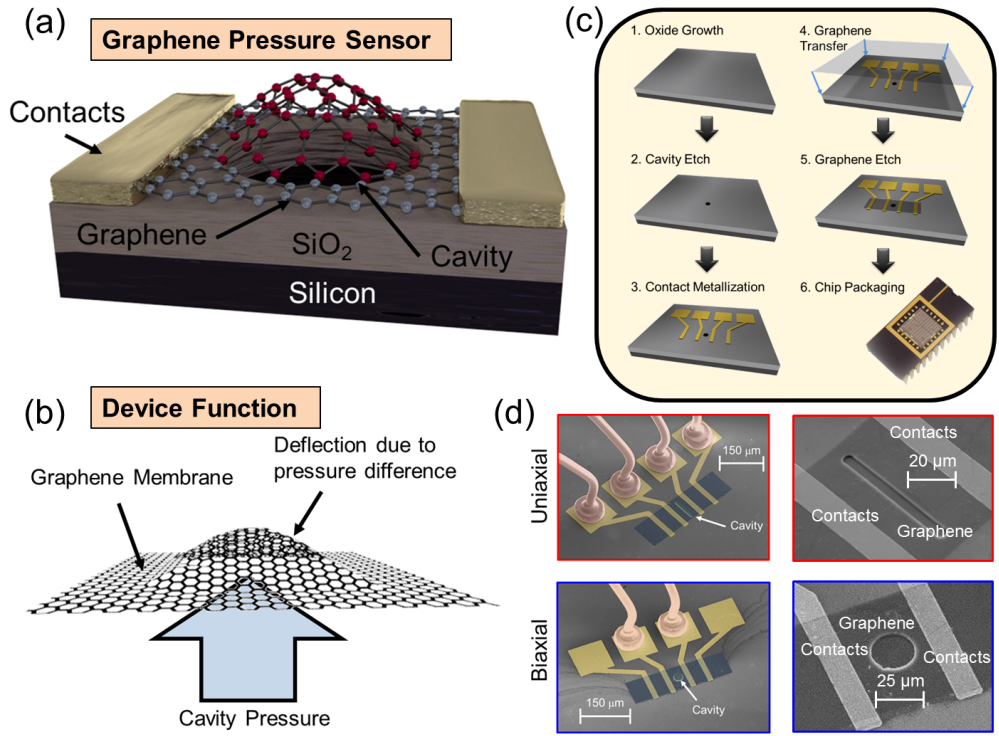

**Figure 1: a)** Schematic of the pressure sensor used in this work. The red area represents the active area of the device. **b)** Representation of membrane functionality in a graphene pressure sensor. As the pressure outside the cavity varies, it causes a deflection and straining of the graphene membrane, thereby changing its electronic properties. **c)** Fabrication process flow starting with SiO₂ growth on a silicon substrate followed by RIE cavity etching. Metal contacts are then patterned followed by the transfer of graphene. The graphene is patterned using a mask in combination with O₂ plasma etching. Finally, devices are wire bonded and placed into a chip package. **d)** Color-enhanced SEM of a sensor device with a rectangular graphene membrane resulting in uniaxial membrane strain (upper image) and SEM of a sensor device with a circular graphene membrane resulting in biaxial membrane strain (lower image). In the SEMs the graphene is shaded in blue, the cavity in green, the electrodes and contact pads in yellow, and the bond wires in orange. To the right of each color enhanced SEM is an SEM showing a close-up of the cavity region for the corresponding devices.

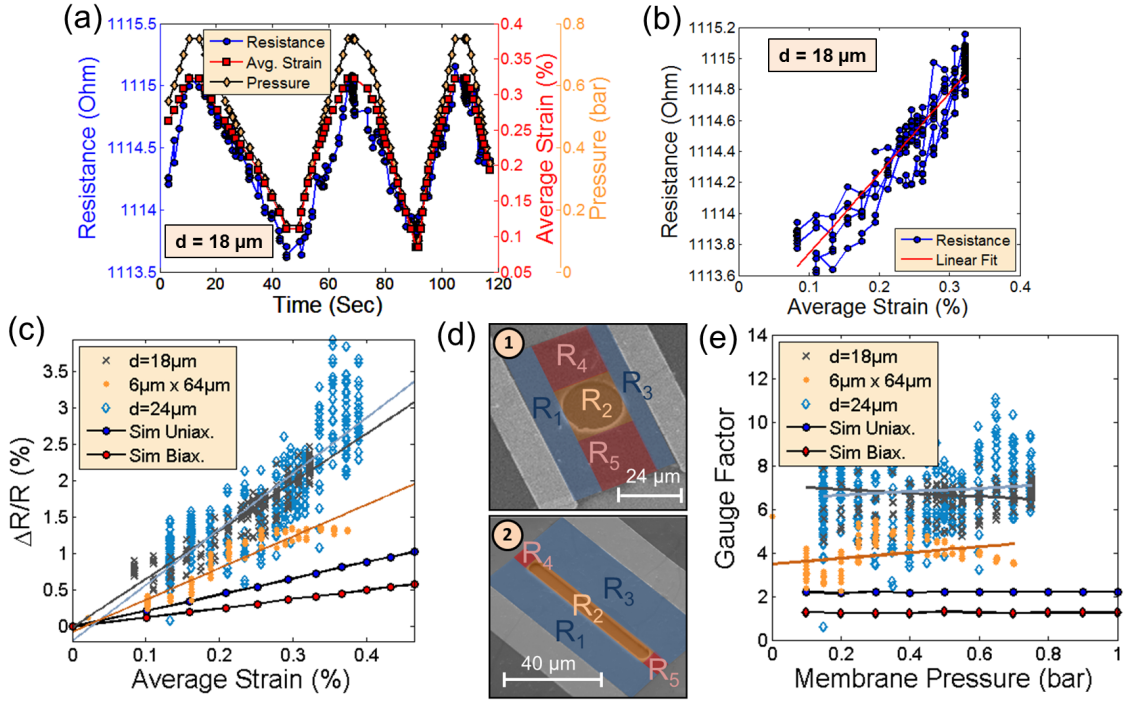

Figure 2: a) Total device resistance, average strain, and vacuum chamber pressure *versus* time for a sensor device consisting of an 18 µm diameter graphene membrane. The resistance and pressure are taken from direct electrical readout, while the average strain is estimated with the help of a strain-deflection COMSOL FEA simulation. b) Resistance *versus* strain relationship for the same device from a). There is a linear relationship between changes in resistance and strain which is predicted by linearized Boltzmann transport simulations. c) Percentage change in resistance of the graphene membrane area for 3 devices with different membrane areas as 18 µm diameter circular membrane (grey x), 24 µm diameter membrane (blue diamonds), and 6 µm x 64 µm rectangular membrane (orange dots). Results from corresponding simulations of the percentage resistance change are shown for uniaxial strain (red), corresponding to the rectangular membrane, and for biaxial strain (blue) corresponding to a circular membrane. d) Color coded areas of the resistors used for the equivalent resistance model for a device with a circular membrane area (1) and rectangular membrane area (2). e) Extracted gauge factors of the different devices compared with the simulation results for both uniaxial and biaxial strained graphene membranes. Note that the gauge factors are constant, regardless of the membrane pressure and the membrane diameter for the circular membranes.

**Table 1: Comparison of gauge factors reported in this work to previous literature.**

|  | Strain Type | Gauge Factor |
|---|---|---|
| **This work** | | |
| Measured | Uniaxial | 3.91 |
| Measured | Biaxial | 6.73 |
| Simulated | Uniaxial | 2.2 |
| Simulated | Biaxial | 1.25 |
| **Previous Literature** | | |
| Zhu[9] | Biaxial | 1.6 |
| Huang[10] | Uniaxial | 1.9 |
| Lee[11] | Uniaxial | 6.1 |
| Wang[12] | Uniaxial | 2 |

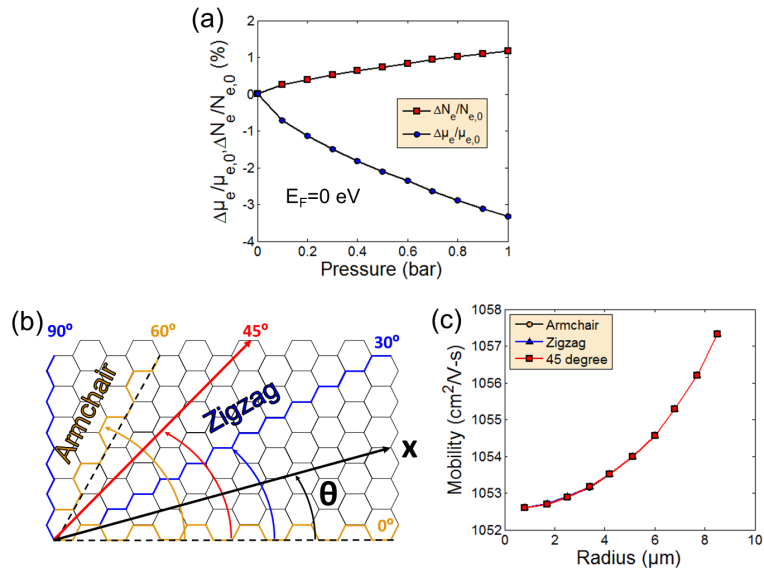

**Figure 3: a)** Simulated percentage change in electron density and mobility *versus* the differential pressure pushing against a graphene membrane. **b)** Schematic of different strain orientations, explored to determine whether orientation of the graphene influences the resistance change. **c)** Electron mobility in a graphene membrane as a function of the membrane radius. Note that the mobility decreases toward the center of the membrane where strain levels are at a maximum.

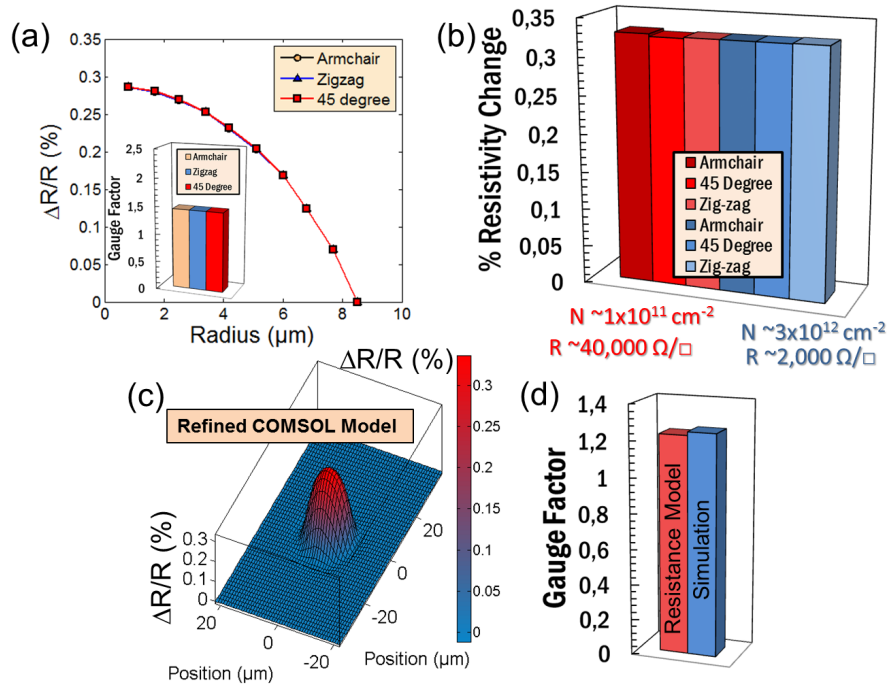

Figure 4: a) Membrane resistance change (ΔR/R) *versus* radius of graphene membranes. The strain model captures how the strain increases towards the center of a membrane (radius of 0 μm). This was performed for armchair (yellow circle), zigzag (blue triangle) and 45 degrees (red square) directions based on Fig. 4c. Virtually no impact of the orientation is observed. Inset: gauge factors extracted for the different orientations. b) Resistivity change for different strain directions for low and high graphene doping (red and blue, respectively). Doping has virtually no effect on resistivity change. c) Simulated map of ΔR/R over the entire graphene patch of the pressure sensor calculated with a refined COMSOL model. d) Comparison of gauge factors extracted from the simulation in c) using the simplified resistance model, described in the supporting information, (red) compared to the gauge factor extracted from charge transport simulations in Fig. 2d. Note that the simplified resistance model provides very close gauge factor estimation supporting the conclusion that the simplified resistance model gives a very good approximation of the gauge factor.

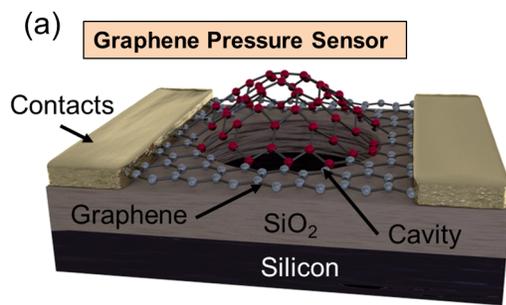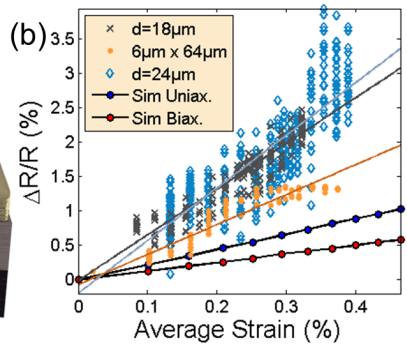

**For Table of Contents Only**